# Coupled superconductors and beyond

## Brian D. Josephson

*Cavendish Laboratory, J.J. Thomson Ave., Cambridge CB3 0HE, UK*
http://www.tcm.phy.cam.ac.uk/~bdj10/

This paper describes the events leading to the discovery of coupled superconductors, the author's move in the 1970s to a perspective where mind plays a role comparable to matter, and the remarkable hostility sometimes encountered by those who venture into unconventional areas.

### Learning about superconductivity

My official Ph.D. project was experimental, not theoretical [1], but having theoretical inclinations I was encouraged by Professor Shoenberg and other members of the low temperature group to study the theoretical aspects of superconductivity. I puzzled over the question 'how do superconductors work?'. The idea that superconductors have a phase was apparent in a number of treatments of superconductivity, starting off with the phenomenological theory of Ginzburg and Landau [2], later justified on the basis of a Green's function treatment by Gor'kov [3]. It was apparent also in the Bogoliubov treatment of superconductivity [4] and the Anderson pseudospin approach [5] which displayed the degree of freedom associated with the phase in graphic form. I recognised that the phase gradient, in accord with the Ginzburg–Landau equations, 'told electrons which way to flow', and that this might happen even in equilibrium. And in the case of a ring, the phase change round a ring would be quantised, leading to the quantised flux observed at about that time by Deaver and Fairbank and by Doll and Näbauer, and implicit in the Ginzburg–Landau theory.

My interest in junctions stemmed from a question put by my supervisor, Brian Pippard, who was sceptical of Giaever's theory for the current through a junction between superconductors [6]. Why, he wondered, did coherence factors not enter into the result as they do for many other phenomena in superconductors? I could see that a possible answer was that the coherence factors for a tunnelling quasiparticle would depend on the difference between the phases on the two sides of the junction, and if these varied the coherence factors might average out to unity. This however raised in my mind the possibility that the phase might be something physical. Symmetry considerations ruled out the possibility of the absolute phase being physical, but not the phase difference between two superconducting regions that can exchange electrons. The next development was Phil Anderson, who was on sabbatical at the Cavendish at the time, showing me a calculation published in Physical Review Letters [7] justifying Giaevar's result, but only in the case where one side was normal, not the more interesting case of two superconductors. I learned later that Falicov had done the same calculation that I did subsequently but was baffled by the extra terms, so the authors decided not to include the two-superconductor case in the paper.

The paper of Cohen *et al.* had treated tunnelling by simply adding to the Hamiltonian terms that transferred electrons across the barrier. I applied their method to the two-superconductor case and got the additional coherence factor terms that I expected, which I thought might manifest as an oscillatory component to the tunnelling current. There seemed to be something wrong, however, as the perturbation calculation produced additional terms that did not vanish at zero applied voltage and implied a supercurrent. I had in fact anticipated a supercurrent but of very small magnitude since the probability of a pair current was expected to be very small compared with the normal current. But my calculation was in fact correct, and the large supercurrent subsequently explained in terms of coherence.

My prediction of tunnelling superconductors, including predictions of ac supercurrents and the magnetic field dependence of the critical current, was published in Physics Letters [8]. It was nine months before the existence of coupled superconductors, and their dependence on magnetic fields, was confirmed by Anderson and Rowell [9]. Later, the anticipated ac supercurrents were observed indirectly by Giaever [10] and later directly by Yanson *et al.* [11].

## New interests: mental phenomena and mind–matter unification

Since the 1970s I have been concerned chiefly with two issues, the problem of the organisation of the mind [12], and what I have named 'Mind–Matter Unification'. The latter stems from the intuition that the role of mind is not fully addressed by conventional theories, and that new physics is sometimes involved. Proposals of this general nature have been made by a number of physicists in the past: for example, Bohr [13] argued that the application of quantum mechanics to life could be problematic, while Wigner and others [14] suggest that consciousness needs to be included in physics to get a fully comprehensive account of nature. These issues I have discussed myself, in various publications [15,16]. A more recent paper [17] develops the idea of Wheeler [18] that 'acts of observer-participancy' are what determine the nature of reality. My paper begins with the not unreasonable proposal that observers be viewed from the standpoint of biology rather than physics. Earlier, in an excursion into the realm of the arts, I collaborated with a musicologist to argue that musical aesthetics points towards specific musical patterns possessing a 'generative capacity' that cannot be understood in conventional terms [19].

A general theme in all this is the idea that biology is 'a different game'. How precisely that game is played is an issue for the future, and there are various directions that we are exploring. My collaborator Fotini Pallikari has illu-strated the situation we seem to be in with the cartoon shown below. The diagram illustrates the fact that the scientist is confronted with a 'hail' of data and candidate theories, and out of these has to try to select the theory that fits the data best. Such a situation led us in the past from classical mechanics to quantum mechanics, and now appears to be leading us to a picture where mind plays a key role.

### Where progress and politics collide

My transition into believing that mind has to be taken seriously as an entity in its own right proved also to be a transition into an environment that was hostile where previously it had been very supportive. The scientific community has its own belief systems that it is dangerous to challenge (consider the case of the winner of the most recent Nobel Prize in Chemistry, Daniel Shechtman, who suffered years of ridicule and hostility from colleagues and friends because of his suggestion that crystals could have aperiodic structures, which should not have been controversial). Being a Nobel Laureate protects one from the worst pressures, but not from curiosities such as this letter relating to a conference to which I had previously been given an invitation and even been asked how long I wished to speak:

*"It has come to my attention that one of your principal research interests is the paranormal ... in my view, it would not be appropriate for someone with such research interests to attend a scientific conference."*

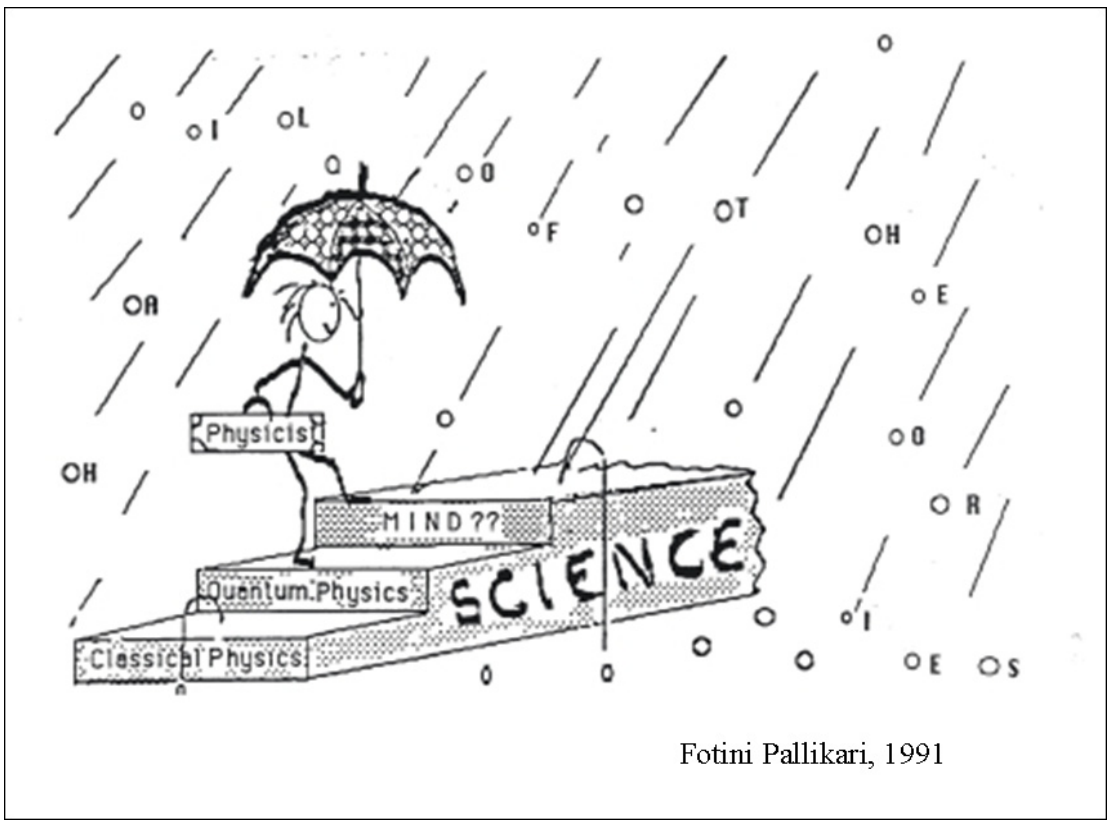

Fotini Pallikari, 1991

I learned from subsequent correspondence that it was feared that my very presence at the meeting might damage the career prospects of students who attended, even if I did not touch on the paranormal in my talk. One is distinctly reminded of Orwell's concept of 'thoughtcrime'!

More seriously, my interest in such matters seems to have led to the harassment of students working with me, even in regard to projects not related to the paranormal. A student who had been offered funding by the laboratory, and was very interested in doing a project examining parallels between classical organisation such as flocking behaviour and quantum wholeness, was told that the funding that had been offered would not be available for a project under my direction. Again, a student who had done a successful computer simulation of development based on the hyperstructure model of Baas [20] was pressured by the department into stopping work on that project on the grounds of it 'not being physics', and had to start afresh on another project. I had hoped that Osborne's programming skills would herald a transition to a firmer basis for my speculative ideas on the organisation of the mind, but it was not to be.

Studying developmental processes on the basis of a different kind of model, that of the neural network, is an accepted research topic for physicists, and one can only marvel at the way the novelty of the picture used in Osborne's simulation provided sufficient grounds for blocking that project. One wonders how much the advance of science in general suffers from such small-minded thinking. All one can say about this [21] is 'it has always been thus'. Some ideas are irrationally perceived as dangerous, and protective mechanisms, usually involving arguments that would fall apart under close examination, are brought up to avoid confronting the possibility that they may be of value.

My original assumption that scientists, being intelligent people, would have the ability to view experimental evidence and theoretical arguments objectively has been severely challenged by my experiences over decades of working in frontier areas of science (a very well known scientist retreated rapidly into the distance, rather than showing interest, when I told him we had an argument [22] that could reconcile quantum mechanics and paranormal phenomena). But, in the end, truth will prevail.

## Acknowledgements

I wish to thank Judith Driscoll and Fotini Pallikari for suggestions concerning the manuscript. A video uploaded by Kelly Neill was the source of the section title "Where progress and politics collide".